\documentclass{article}

\usepackage{PRIMEarxiv}

\usepackage[utf8]{inputenc} 
\usepackage[T1]{fontenc}    
\usepackage{hyperref}       
\usepackage{url}            
\usepackage{booktabs}       
\usepackage{amsfonts}       
\usepackage{nicefrac}       
\usepackage{microtype}      
\usepackage{lipsum}
\usepackage{fancyhdr}       
\usepackage{graphicx}       
\graphicspath{{media/}}     

\title{Software and Security Engineering in Digital Transformation}

\author{
  Mamdouh Alenezi \\
  College of Computer and Information Sciences \\
  Prince Sultan University \\
  Riyadh, Saudi Arabia\\
  \texttt{malenezi@psu.edu.sa} \\
}

\begin{document}
\maketitle

\begin{abstract}
Digital transformation is a hot topic in the current global environment as a large number of organizations have been working to adopt digital solutions. Software engineering has also emerged to be a more important role as a large number of systems, either traditional or smart, are dependent on the software that collects, store, and process data. The role of software engineers has also become crucial in digital transformation. In this regard, this paper aims to examine the trends of software engineering and the role of software engineers in digital transformation. In addition to this, this paper also examines the importance of secure software development in digital transformation. It can be concluded that software engineering is an integral part of digital transformation as all digital systems make use of software to perform their functions efficiently. Software act as a bridge between digital systems and humans to use the systems interactively and efficiently.
\end{abstract}


\section{Introduction}

\begin{figure}[htbp]
\begin{center}
\includegraphics[scale=1]{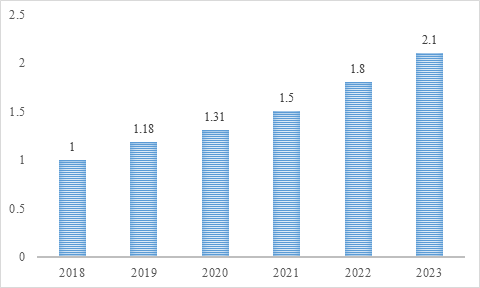}
\end{center}
\caption{Spending on digital transformation worldwide from 2018 to 2023 (in trillion U.S. dollars).\label{fig44}}
\end{figure}

Digital transformation can be seen from the perspective of the link between the structural, strategic, and technological changes that are imperative to meet the demands of the contemporary digital era \cite{drechsler2020crossroads}. Digital transformation, led by organizational strategy, is causing explosive growth in digital organizations \cite{kane2015strategy}. It is creating new ways to engage customers, collaborate with partners, and achieve operational efficiency. Digital transformation is gaining a lot of momentum in the last few years. Figure \ref{fig44} shows the trending increase of spending worldwide from 2018 to 2023 as reported by statista.com. This indicates that digital transformation is becoming a reality throughout sectors.

\begin{figure}[htbp]
\begin{center}
\includegraphics[scale=0.6]{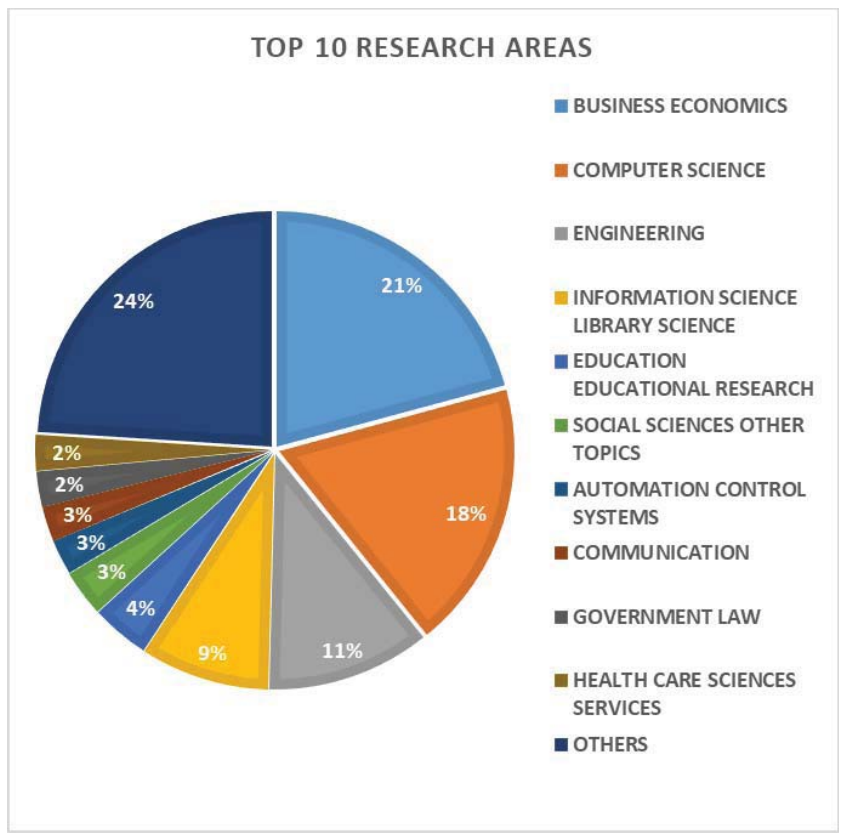}
\end{center}
\caption{Research and Development Areas in Digital Transformation.\label{fig11}}
\end{figure}

Digital transformation is a hot topic in the current global environment as a large number of organizations have been working to adopt digital solutions. As per the study of Sousa and Rocha \cite{sousa2019digital}, technologies like artificial intelligence and internet-of-things (IoT) have gained high popularity due to their ability to automate the systems and make the existing procedures smarter. Thus, these advanced systems are being employed increasingly in various areas like traffic management, healthcare systems, and other organizational functions. In this regard, software engineering has also emerged to be a more important role as a large number of systems, either traditional or smart, are dependent on the software that collects, store, and process data \cite{kutnjak2019digital}. Figure \ref{fig11} shows that digital transformation has entered almost every sector, which also shows the high potential of software engineering. Therefore, the role of software engineers has also become crucial in digital transformation. In addition to this, digital transformation has engaged various complex software development practices and activities to develop the most advanced systems that can solve the existing social and economic problems \cite{ebert2016requirements}. Hence, it can be noted that software engineering is key to an effective and efficient digital transformation globally. In this regard, this paper aims to examine the trends of software engineering and the role of software engineers in digital transformation. In addition to this, this paper also examines the importance of secure software development in digital transformation.

\section{Software Engineering in Digital Transformation}

\begin{figure}[htbp]
\begin{center}
\includegraphics[scale=0.4]{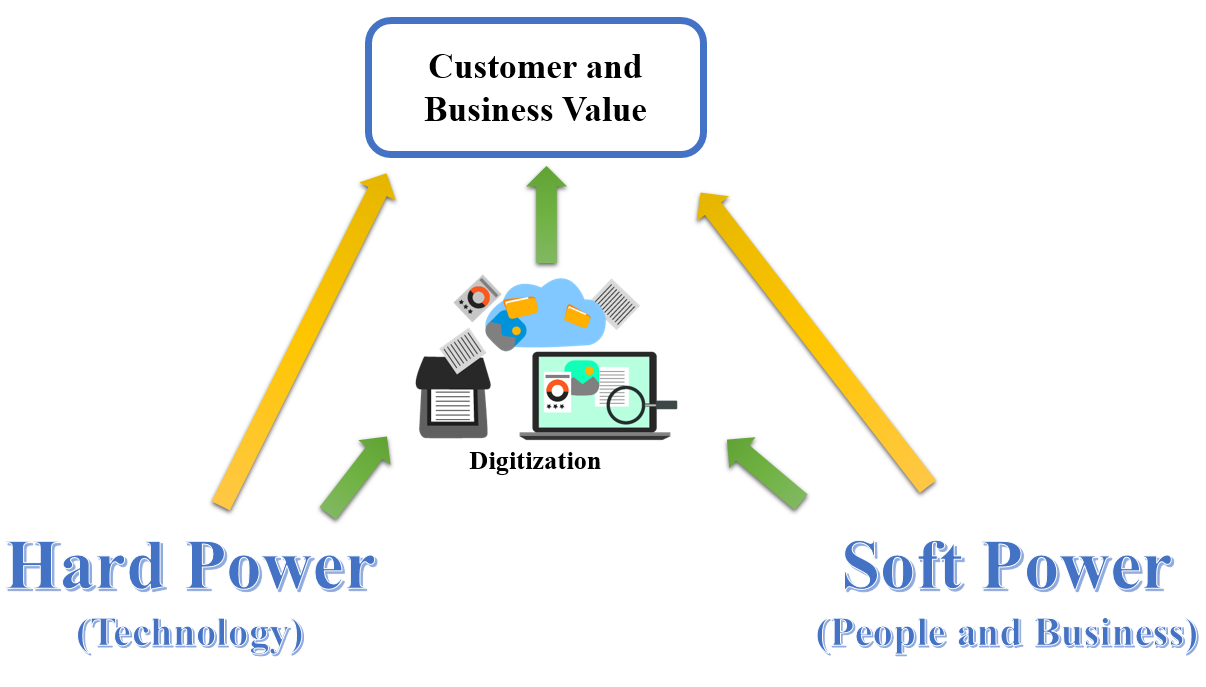}
\end{center}
\caption{Combining Soft Power and Hard Power Through Digitization.\label{fig22}}
\end{figure}

Digital transformation is a broad concept that aims to foster positive development in social and economic aspects globally \cite{baker2015digital}. Various organizations have been operating on traditional methods and systems that limit their ability to show high efficiency. Traditional systems are not digital that causes them to require human support that adds to their inaccuracies and inefficiencies. Therefore, digitization of the systems has become a major need of time due to the growing business needs of the organizations. In this regard, software engineering has a major role to play as it sets the base of the systems that are used by the organizations. As per Ulas \cite{ulas2019digital}, software engineering refers to the design, development and implementation of software that can work on different devices like PC, tablets, and smartphones. These software can run on the web (web applications) or can be downloaded on devices \cite{scuotto2021microfoundational}. Thus, through the use of these software, users can perform the desired function for which these applications are designed. All the organizations in the current world make use of software in any of their functions to enhance their operations. Ebert and Duarte \cite{ebert2018digital} reveal that digitization of the systems offers a link between the hard and soft technology that helps the organizations achieve enhanced customer and business value (see Figure \ref{fig22}).

\begin{figure}[htbp]
\begin{center}
\includegraphics[scale=1]{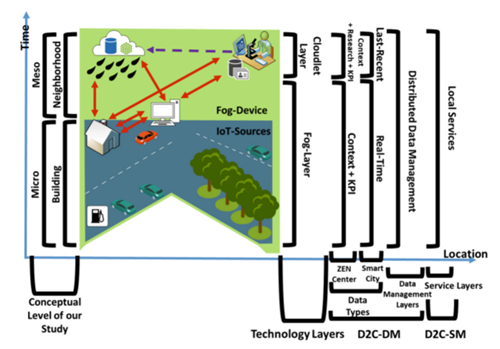}
\end{center}
\caption{An IoT System for Weather.\label{fig33}}
\end{figure}

Thus, it can be noted that software engineering holds immense importance in the digital transformation of organizations. Besides this, software and applications are being increasingly used by the public. The major aim of these digital apps is to provide convenient and efficient services to users \cite{li2018digital}. For instance, e-commerce apps have become highly popular in recent years as they have provided people with the solution to shopping for products at their choice of time and place. Therefore, such software solves the problems that previously existed for the people \cite{reinartz2019impact}. It is important to note that software engineering plays a crucial part in the development of these applications. Software engineering involves various processes like determining the business needs, identifying the customers’ requirements, designing the layout of the software, and developing the final application \cite{karpunina2019economic}. Based on these processes, it can be noted that software engineering is an integral part of digital transformation as all the digital applications and systems that are either used by an individual or an organization make use of software to accomplish the desired goals. Besides this, advanced technologies, like IoT, big data analytics, simulation and modeling, and digital twin, require effective and efficient software that can take commands from the user, process them, and display the output \cite{pappas2019digital}. Software engineering leads to the development of user-friendly systems that can be easily used by people so that they can interact with technology successfully. Otherwise, the language used by the computer is not comprehendible by humans, which would have left the benefits of these technologies in vain. Figure \ref{fig33} shows a system that is based on IoT for weather forecast; the usability of this system is majorly dependent on the software. Therefore, software plays a critical role in being a bridge between humans and technologies to interact \cite{hess2016options}. Hence, it can be stated based on these findings that digital transformation would become highly difficult and almost impossible if software engineering is removed from the process.

Moreover, digital transformation comprises various areas of development that include cloud computing applications, machine learning, and augmented reality systems. These digital developments are highly effective in automating the processes and providing improved experiences to the people \cite{frankiewicz2020digital}. Software engineering plays a significant role in making these technologies are reality as all these systems rely on software to be used by the people. Similarly, digital transformation in other areas like communication is dependent on the use of software that enables people to interact with the digital systems. In addition to this, the world is experiencing major changes due to the rise of technology, which has increased the number of smartphone users, the need for more flexibility, and diversity in technologies \cite{mandviwalla2021small}. Software engineering effectively handles these challenges as the developed software can provide a one-stop solution for the users to access the systems on their smartphones, allowing higher flexibility and integration of different technologies. Therefore, digital transformation is significantly eased through a solid role played by software engineering.

\section{Software Engineers in Digital Transformation Era}

Software engineers are focusing on multiple domains in the digital transformation era. Unlike other professions, software engineers have made their presence as leaders in the technology industry to bring innovation in the digital transformation process. One of the fundamental roles played by software engineers is designing such software and digital products that are innovative in terms of their features and usage. The other focus of the software engineers has been on business sustainability through eco-friendly means of productions \cite{elsafty2021new}. Managing cyber security and developing such techniques that can protect the system from hackers and attackers is another incredible role that software engineers are playing in the digital transformation era. It is an outcome of the sheer hardware and efforts made by software engineers that industry has now moved from legacy monolithic systems towards the flexible digital foundation. Another role is bringing customer and user experience (CX/UX) in the technology-centric model that minimizes any barriers between the technology company and the client \cite{ardito2012software}. This has resulted in enhanced growth in the IT sector due to the increased digital transformation during the pandemic \cite{bworld}. Moreover, strong technical knowledge is also increasing in the IT sector, such as KMS Technology Vietnam. The company has employed its software engineers to bring digital transformation to the market. Consequently, more than 1000 employees have started services in the IT sector of the Vietnam region, contributing to the success of the sector. 

To fulfil this role, companies are now hiring test engineers so that software engineers, along with their team, can meet the expectations of the client \cite{KMSworld}. Software engineers also play a role in minimizing the bugs, making changes in the ongoing transformation process through scrum meetings. In other words, software engineers have minimized the stress in the IT industry by breaking down the product development process into several chunks. As a result, the greater productivity in the IT sector is an ultimate outcome of the software engineers and their innovative approach \cite{singh2017chief}. Furthermore, the software engineers are now playing their role to bring problem-solving techniques and solutions that only introduce new features, updates and innovativeness in the system. Software engineers have also introduced the IT sector towards an agile project management approach with communicational skills that contribute to the success of the project. Google, Amazon, LinkedIn have made leading market positions due to their full-stack software developers such as JavaScript, programming languages, and even developing big data analytics and machine learning techniques \cite{vander2012shipping}. Moreover, software engineers have also offered a different perspective in the IT sector that soft skills are important when implementing the solution. This is why software engineers are now considered advocates who can alter the solutions as per the nature, complexity and technicality of the system. Finally, software developers have proven that IT leaders are the emerging need of the industry as the developers also need to inculcate transformational leadership skills to bring innovation \cite{li2020distinguishes}. Thus, software engineers have changed the dynamics of the software industry in the digital transformation era.

The more we connect services with our lives, the more we need software engineering excellence, ranging from specifying requirements to developing and continuously deploying reliable, safe, and secure software. Dependable business models demand software security, maintainability, and sustainability. Excellent software engineers model business processes, functionality, and architecture from a systems perspective while ensuring robustness and security. The hierarchic modeling of business processes, functionality, and architecture from a systems perspective allows for early simulation while ensuring robustness and security.

\section{Importance of Secure Software in Digital Transformation}

However, it must also be noted that the rapid digital transformation has also caused an increase in the number of security threats. Cybersecurity is recognized as a substantial cross-cutting concern that influences various aspects of digital transformation, from the choice of technology to the financial outcomes \cite{teoh2017national}. Cybercriminals have become highly active and are increasing with time, which has become a major threat to the integrity of digital systems. Due to this, it is highly important for the software to be secure so that they are not breached by the attackers and can keep the data of the users safe. As per the study of Duc and Chirumamilla \cite{duc2019identifying}, attackers often look for loopholes in the designs and architectures of software to gain access to confidential data of an individual or organization. Therefore, during software development, it is essential that security aspects are considered at all the stages so that an effective and strong software is engineered that has the ability to counter all the security threats. However, threats like malware, DoS attacks, and hacking are still persistent and continue to affect people. Nonetheless, it was revealed by Doukidis et al. \cite{doukidis2020digital} that the use of effective security protocols in software could enhance the ability of the systems to show high resistance against the threats and reduce the chances of any potential breach.

Secure software holds major importance for digital transformation as digital systems store, process, and transfer the data of the people through electronic means. Unsecure software can be a major threat to the digital transformation of society and organizations as it would make the confidential data of people vulnerable to various security threats. It was reported by Collett \cite{ciso1} that software security is the second-highest priority of investors in digital transformation, which shows the urge for organizations to have secured systems. The experts are wary of the fact that the rising cybersecurity issues can cause hindrance in rapid and effective digital transformation. As software make a major part of digital transformation, it is one of the major goals of the experts to ensure that the best and most secure software is developed \cite{mendhurwar2021integration}. Through this practice, the digital transformation can be saved from the threats of data leakage, privacy breach, or financial fraud. Hence, software security has a major role in ensuring a swift and efficient digital transformation globally.

\section{Conclusion}

In the end, it can be concluded that software engineering is an integral part of digital transformation as all digital systems make use of software to perform their functions efficiently. Software act as a bridge between digital systems and humans to use the systems interactively and efficiently. Various organizations are aiming for digital transformation so that they can enhance their operations. In this regard, the design, development, and implementation of effective software is essential. Due to the high significance of software engineering, software engineers have a major role to play. These professionals are involved in various practices like design, development, implementation, and testing of the software to align it with the digital needs of an organization. Therefore, software engineers have an important role in digital transformation as they understand the needs of the firms and develop software and systems to meet their requirements. However, digital transformation is affected by the cybersecurity challenges that hinder the effective operations of organizational systems. Unsecure software can lead to data loss and theft, which can cause massive financial losses and a bad reputation. Hence, experts have been highly focused on the development of secure software to ensure that the digital transformation is smooth and effective.

\bibliographystyle{unsrt}  
\bibliography{references}

\begin{thebibliography}{10}

\bibitem{drechsler2020crossroads}
Katharina Drechsler, Robert Gregory, Heinz-Theo Wagner, and Sanja Tumbas.
\newblock At the crossroads between digital innovation and digital
  transformation.
\newblock {\em Communications of the Association for Information Systems},
  47(1):23, 2020.

\bibitem{kane2015strategy}
Gerald~C Kane, Doug Palmer, Anh~Nguyen Phillips, David Kiron, Natasha Buckley,
  et~al.
\newblock Strategy, not technology, drives digital transformation.
\newblock {\em MIT Sloan Management Review and Deloitte University Press},
  14(1-25), 2015.

\bibitem{sousa2019digital}
Maria~Jos{\'e} Sousa and {\'A}lvaro Rocha.
\newblock Digital learning: Developing skills for digital transformation of
  organizations.
\newblock {\em Future Generation Computer Systems}, 91:327--334, 2019.

\bibitem{kutnjak2019digital}
A~Kutnjak, I~Pihiri, and M~Tomi{\v{c}}i{\'c} Furjan.
\newblock Digital transformation case studies across industries--literature
  review.
\newblock In {\em 2019 42nd International Convention on Information and
  Communication Technology, Electronics and Microelectronics (MIPRO)}, pages
  1293--1298. IEEE, 2019.

\bibitem{ebert2016requirements}
Christof Ebert and Carlos Henrique~C Duarte.
\newblock Requirements engineering for the digital transformation: Industry
  panel.
\newblock In {\em 2016 IEEE 24th International Requirements Engineering
  Conference (RE)}, pages 4--5. IEEE, 2016.

\bibitem{baker2015digital}
Mark Baker.
\newblock {\em Digital transformation}.
\newblock Buckingham Business Monographs, 2015.

\bibitem{ulas2019digital}
Dilber Ulas.
\newblock Digital transformation process and smes.
\newblock {\em Procedia Computer Science}, 158:662--671, 2019.

\bibitem{scuotto2021microfoundational}
V~Scuotto, M~Nicotra, M~Del~Giudice, N~Krueger, and GL~Gregori.
\newblock A microfoundational perspective on smes’ growth in the digital
  transformation era.
\newblock {\em Journal of Business Research}, 129:382--392, 2021.

\bibitem{ebert2018digital}
Christof Ebert and Carlos Henrique~C Duarte.
\newblock Digital transformation.
\newblock {\em IEEE Softw.}, 35(4):16--21, 2018.

\bibitem{li2018digital}
Liang Li, Fang Su, Wei Zhang, and Ji-Ye Mao.
\newblock Digital transformation by sme entrepreneurs: A capability
  perspective.
\newblock {\em Information Systems Journal}, 28(6):1129--1157, 2018.

\bibitem{reinartz2019impact}
Werner Reinartz, Nico Wiegand, and Monika Imschloss.
\newblock The impact of digital transformation on the retailing value chain.
\newblock {\em International Journal of Research in Marketing}, 36(3):350--366,
  2019.

\bibitem{karpunina2019economic}
Evgeniya~K Karpunina, Maria~E Konovalova, Julia~V Shurchkova, Ekaterina~A
  Isaeva, and Alexander~A Abalakin.
\newblock Economic security of businesses as the determinant of digital
  transformation strategy.
\newblock In {\em Institute of Scientific Communications Conference}, pages
  251--260. Springer, 2019.

\bibitem{pappas2019digital}
Ilias~O Pappas, Patrick Mikalef, Yogesh~K Dwivedi, Letizia Jaccheri, John
  Krogstie, and Matti M{\"a}ntym{\"a}ki.
\newblock Digital transformation for a sustainable society in the 21st century.
\newblock In {\em 18th IFIP WG 6.11 Conference on e-Business, e-Services, and
  e-Society, I3E 2019, Trondheim, Norway, September 18--20, 2019}, page~20.
  Springer, 2019.

\bibitem{hess2016options}
Thomas Hess, Christian Matt, Alexander Benlian, and Florian Wiesb{\"o}ck.
\newblock Options for formulating a digital transformation strategy.
\newblock {\em MIS Quarterly Executive}, 15(2), 2016.

\bibitem{frankiewicz2020digital}
Becky Frankiewicz and Tomas Chamorro-Premuzic.
\newblock Digital transformation is about talent, not technology.
\newblock {\em Harvard Business Review}, 6:3, 2020.

\bibitem{mandviwalla2021small}
Munir Mandviwalla and Richard Flanagan.
\newblock Small business digital transformation in the context of the pandemic.
\newblock {\em European Journal of Information Systems}, pages 1--17, 2021.

\bibitem{elsafty2021new}
Ashraf Elsafty, Ahmed Elzeftawy, et~al.
\newblock The new era of digital transformation and covid-19 effect on the
  employment in mobile operators in egypt.
\newblock {\em Business and Management Studies}, 7(1):74--99, 2021.

\bibitem{ardito2012software}
C~Ardito, P~Buono, MF~Costabile, and R~Lanzilotti.
\newblock Are software companies aware of ux?
\newblock In {\em International Workshop on the Interplay between User
  Experience (UX) Evaluation and System Development (I-UxSED 2012)}.

\bibitem{bworld}
Kimberly Mlitz.
\newblock {BWorld Robot Control Software}.
\newblock
  \url{https://www.statista.com/statistics/1200484/covid-digital-transformation-process-industry/},
  2021.
\newblock [Online; accessed 09-December-2021].

\bibitem{KMSworld}
KMS Technology.
\newblock {Industry practitioners share: Software engineers in the digital
  transformation era}.
\newblock
  \url{https://kms-technology.com/press/software-engineers-digital-transformation-era},
  2021.
\newblock [Online; accessed 09-December-2021].

\bibitem{singh2017chief}
Anna Singh and Thomas Hess.
\newblock How chief digital officers promote the digital transformation of
  their companies.
\newblock {\em MIS Quarterly Executive}, 16(1), 2017.

\bibitem{vander2012shipping}
Chris Vander~Mey.
\newblock {\em Shipping Greatness: Practical lessons on building and launching
  outstanding software, learned on the job at Google and Amazon}.
\newblock "O'Reilly Media, Inc.", 2012.

\bibitem{li2020distinguishes}
Paul~Luo Li, Amy~J Ko, and Andrew Begel.
\newblock What distinguishes great software engineers?
\newblock {\em Empirical Software Engineering}, 25(1):322--352, 2020.

\bibitem{teoh2017national}
Chooi~Shi Teoh and Ahmad~Kamil Mahmood.
\newblock National cyber security strategies for digital economy.
\newblock In {\em 2017 International Conference on Research and Innovation in
  Information Systems (ICRIIS)}, pages 1--6. IEEE, 2017.

\bibitem{duc2019identifying}
Anh~Nguyen Duc and Aparna Chirumamilla.
\newblock Identifying security risks of digital transformation-an engineering
  perspective.
\newblock In {\em Conference on e-Business, e-Services and e-Society}, pages
  677--688. Springer, 2019.

\bibitem{doukidis2020digital}
Georgios Doukidis, Diomidis Spinellis, and Christof Ebert.
\newblock Digital transformation-a primer for practitioners.
\newblock {\em IEEE Software}, 37(5):13--21, 2020.

\bibitem{ciso1}
Stacy Collett.
\newblock {What is security's role in digital transformation?}
\newblock
  \url{https://www.csoonline.com/article/3512578/what-is-securitys-role-in-digital-transformation.html},
  2020.
\newblock [Online; accessed 09-December-2021].

\bibitem{mendhurwar2021integration}
Subodh Mendhurwar and Rajhans Mishra.
\newblock Integration of social and iot technologies: architectural framework
  for digital transformation and cyber security challenges.
\newblock {\em Enterprise Information Systems}, 15(4):565--584, 2021.

\end{thebibliography}

\end{document}